\documentstyle[amsfonts,epsf,aps]{revtex}

\newcommand{\ee}{\end{equation}}
\newcommand{\be}{\begin{equation}}
\begin{document}
\title{
First-order $\Theta$-point of a single polymer chain}
\author{Annick  Lesne and Jean-Marc Victor} 
  
\address{
Laboratoire de Physique Th\'eorique 
des Liquides, \\
Universit\'e Pierre et Marie Curie,  \\
Case courrier 121, 4 Place Jussieu, 75252
 Paris Cedex 05,
France\\
}
\date{\today}
\maketitle

\begin{abstract}
{\small

Conformational transitions of a single
macromolecule of finite size $N$ cannot be described within  standard
thermodynamic framework.
Taking as a basis  a simple model 
of homopolymer exhibiting a coil-globule transition,
we show that a relevant approach is to 
describe the thermal equilibrium distribution $P_N^{(\beta)}(t)$
of some  variable $t$ characterizing the conformation.
Although the  mean order parameter exhibits a second-order behaviour
in the infinite-size limit,
the $\Theta$-point arises from the coexistence of two distinct
populations, associated with two well-separated peaks of $P_N^{(\beta)}(t)$
and identified respectively with a coil state and a globule state.
Remarkably, this first-order feature increases with the size of the chain.
It allows to describe the transition within a two-state model,
well-suited to analyse experimental data.
}
\end{abstract}


\twocolumn
The recent advances of single molecule experiments allow a direct access
to the microscopic conformation of a macromolecule at thermal
equilibrium\cite{Science}.
Interpretation of such experimental data requires to stick to a
finite-size framework. 
In particular, a straightforward application of 
the standard thermodynamic description 
and classification of phase transitions to 
conformational transitions of isolated  macromolecules of size $N$
is highly questionable.
We promote the analysis of the thermal
equilibrium distribution $P_N^{(\beta)}(t)$
of some conformational   parameter $t$ of the macromolecule.
$\beta$ is as usual equal to $1/k_BT$, where $k_B$ is the Boltzmann constant
and $T$ the absolute temperature.
We illustrate such a procedure and its motivations 
on a simple model of 
coil-globule transition.
A main puzzle is the apparent contradiction between the tricritical scaling
properties predicted at the $\Theta$-point\cite{PGG} and the experimental
observations\cite{Yosh}
% (for instance those of Yoshikawa et al.) 
of coexistence of two
populations with well-separated features,  identified with 
%the coexistence of 
coil and globule states. 
%Our aim is to bridge the gap between  statistical description
%in the configurational space and macroscopic properties, in order that theoretical
%results give relevant predictions about observed systems.

\vskip 2mm
In a previous study\cite{ILV}, we evidenced that 
the relevant order parameter
$t$ to unravel the  coil-globule transition of an isolated polymer 
chain of size $N$ with
excluded-volume interactions
is a power of the 
 density $\rho=Nr^{-3}$, where $r$ is 
 the radius of gyration   in a given configuration:
\be
t=\rho^{1/(\nu d-1)}=\left({N/ r^3}\right)^{5/4}
\ee
(here $d=3$ and $\nu=3/5$ is the Flory exponent).
The  distribution $P_N(t)$ at infinite 
temperature
 has been deduced from scaling
arguments  supplemented with numerical simulations.
%, namely 
% Monte Carlo sampling on a cubic lattice.
%It provides the exact entropic contribution
%(coming from excluded-volume effects)
% to the conformational transitions
%of a single self-avoiding walk.
Choosing an energy
$U=-NJt$  where $J$ is some coupling constant, that accounts 
within a mean-field approximation 
for attractive interactions at contacts\cite{JM},
the  
%analytical 
expression $P_N^{(\beta)}(t)$
for the Boltzmann-Gibbs distribution of the chain 
%in $t$-space 
is:
% in $t$-space follows:
\be P_N^{(\beta)}(t)\sim t^c\,e^{-A^{\prime}
(Nt)^{-q}}\,
e^{-N[(A-\beta J)t+Bt^n]}
\ee
$A$, $A^{\prime}$, $B$, $c$ and $n$ are numerical constants fitted on
simulation data, with $n\approx 2$ and 
$c$  undoubtedly lower than $-1$
($c\approx -1.13$).
We  introduce a reduced temperature:
\be
\tau=1-\theta/T=1-\beta J/A\ee
%Indeed, according to the sign of $\tau$, the 
%dominant contribution to
%$P_N^{(\beta)}(t)$ is located in different
% domains of values of $t$.
For $\tau>0$, the 
distribution is strongly
peaked around a value of $t$ of order 
$1/N$.	
This corresponds to values of $r$ of 
order 
$N^{\nu}$; 
%where $\nu=3/5$ is the Flory exponent; 
 it leads to
identify this high temperature regime 
with a {\it coil phase}.
For $\tau<0$,
$P_N^{(\beta)}
(t)$ is now peaked in the region  where $t$
 is of order 1.
This corresponds to values of $r$ of order 
$N^{1/d}$
% ($N^{1/d}$ with $d=3$) 
and leads to
identify this low temperature regime with a
 {\it globule phase}.
The  temperature $\theta=J/k_BA$ 
thus gives a rough estimate of the
transition temperature.

%\vskip 2mm
Focusing on the transition, we 
investigate more precisely
the size and temperature dependence of  the distribution $P_N^{(\beta)}(t)$.
%brings out two remarkable features.
% $P_N^{(\beta)}(t)$ 
It involves a scale-invariant
factor: 
\be h(\hat{\tau},\hat{t})=\hat{t}^ce^{-A\hat{\tau}\hat{t}-B\hat{t}^n}
\ee
where the scaling variables are:
\be
\hat{t}=tN^{1/n}\hspace{10mm}{\rm and}\hspace{10mm}\hat{\tau}=\tau N^{1-1/n}
\ee
The corresponding distribution writes:
\be
\hat{P}_N(\hat{\tau},\hat{t})
=\frac{h(\hat{\tau},\hat{t})\;
e^{-A^{\prime}N^{-q(1-1/n)}\hat{t}^{-q}}}{{\cal I}_c(N,\hat{\tau})}
\ee
where the normalization factor ${\cal I}_c(N,\hat{\tau})$,
strongly depending on the value of $c$ as we shall see, 
ensures that $\int_0^{\infty}\hat{P}_N(\hat{\tau},\hat{t})d\hat{t}=1$.
One might guess at first sight that a
 scaling regime should be obtained 
at fixed values of $\hat{\tau}$ when $N\rightarrow\infty$.
 But due to the value $c<-1$, the limiting
function $h(\hat{\tau},\hat{t})=\hat{t}^ce^{-A\hat{\tau}\hat{t}-B\hat{t}^n}$
 is not integrable in $\hat{t}=0$.
This obviously compels to focus on the finite-size distribution
$\hat{P}_N(\hat{\tau},\hat{t})$: the relevance of the 
size-dependent contribution
factor $e^{-A^{\prime}N^{-q(1-1/n)}\hat{t}^{-q}}$
in  
%the normalization factor
 ${\cal I}_c(N,\hat{\tau})$ breaks the scale invariance.
%We shall  investigate the consequences of this feature
%on the coexistence diagram and tricritical scaling of the moments. 
In order to elucidate the influence of the factor $\hat{t}^c$
on the conformational transition,
 we shall compare the actual case $c\approx -1.13$ with more
general
values of $c$, in particular $c>-1$.

\vskip 2mm
A bimodal shape of $\hat{P}_N(\hat{\tau},\hat{t})$ is never  observed
for $c\geq 0$, neither for $c<0$ and $N$ smaller than
 a value $N_0(c)\sim |c|^{-4}$ 
%(2.92/|c|)^4$
 ($N_0=45$ for $c=-1.13$);
only one peak exists and it slowly shifts from the coil region towards the
globule region as temperature decreases.
In this case, 
represented on Figure 1a, the transition is continuous and no temperature of transition
can be clearly defined. 
\vskip 2mm
Let us now turn, in all what follows, to the case $c<0$.
 For $N\geq N_0(c)$,  $P_N^{(\beta)}(t)$ exhibits two
peaks in some range of temperatures.
Indeed, a 
%$c<-1$ and arbitrarily large $N\geq N_0=45$ 
% and inside a definite interval of 
%rescaled temperatures $[\hat{\tau}_c(N), \hat{\tau}_g(N)]$,
% $P_N^{(\beta)}(t)$ exhibits two marked and well-separated
%peaks.
 globule peak exists for rescaled temperatures
$\hat{\tau}<\hat{\tau}_g(N)$, where $\hat{\tau}_g(N)$
slightly decreases from a critical value $\hat{\tau}_0=
\hat{\tau}_g(N_0)<0$ 
%(here $\hat{\tau}_0\approx -3.2$)
 to an asymptotic value $\hat{\tau}_g(\infty)\sim -|c|^{(n-1)/n}$.
The actual value $c=-1.13$ gives  $\hat{\tau}_0\approx -3.2$
and $\hat{\tau}_g(\infty)\approx -4$.
%\be \hat{\tau}_g(c,\infty)=
%-\,
%{n\;(nB)^{1/n}\over A}\;\left({|c|\over n-1}\right)^{(n-1)/n}\approx -4
%\ee
%Obviously $\hat{\tau}_g(c=0,\infty)=0$.
%On the other side,  the
A coil peak exists as soon as $\tau>\tau_c(N)$ where 
$\tau_c(N)$ rapidly increases
from $\tau_0$ towards an asymptotic value $\tau_c(\infty)<0$ independent of $N$
(here  $\tau_c(\infty)\approx -0.4$). The rescaled bound 
$|\hat{\tau}_c|$ thus behaves as $N^{1-1/n}$, so
that a coil peak always exists in the scaling region ($\hat{\tau}$
finite) for $N$ enough large.
%Note that $N_0(c)$ tends to $\infty$ as $c$ tends to 0: it would require
%larger and larger sizes to observe coexistence of two peaks if $c$
%is to be increased towards 0.

% becomes arbitrarily large as $N$ increases so that
%any fixed  $\hat{\tau}<0$ will be on the right side of
%$\hat{\tau}_c$ for $N$ enough large.

\vskip 2mm
The bimodal shape of the single molecule distribution reflects
straightforwardly on the statistics describing a dilute solution (enough
dilute to neglect interactions between different chains): it indicates that a
coil population and a globule population coexist
in the  interval of 
rescaled temperatures $[\hat{\tau}_c(N), \hat{\tau}_g(N)]$. 
As represented on Figure 1b, the thermal transition 
at fixed $N$ is achieved through
 an exchange of weight between the two peaks.
A crucial point is that their positions $\hat{t}_c$ and $\hat{t}_g$
remain well-separated when  temperature varies: they are
 located on each side of a
value $\hat{x}(N)$ increasing with $N$ from $\hat{x}_0$
towards and asymptotic value $\hat{x}_{\infty}=[|c|/n(n-1)B]^{1/n}$
(in our case, $\hat{x}_0\approx 13$
and $\hat{x}_{\infty}\approx 26$).
This seems to indicate that our model exhibits a first-order 
%$\Theta$-point.
coil-globule transition\cite{allegra}.
We shall now give a stronger support of this assertion.

\vskip 2mm
A first caveat concerns the experimental reality  of the coexistence.
It is actually possible to distinguish two populations only if the 
height of 
$\hat{P}_N(\hat{\tau},\hat{t})$ at the minimum $\hat{t}_m$
(located between $\hat{t}_c$ and $\hat{t}_g$)
differs significantly from the height of the peaks.
We have checked that it is true for $N$ enough large.
It is then sensible
to partition  the configuration space   in
two disjoint  ``macrostates'':
a coil state $\{\hat{t}<\hat{x}(N)\}$ and
a globule state $\{\hat{t}>\hat{x}(N)\}$
as the peaks remain located on each side of $\hat{x}(N)$ as soon as they exist,
even alone.
Coexistence is actually observed if the  fractions  of molecules
 in each state, i.e.  the areas
of the two  peaks, have comparable values. Let:
\be\label{kappa}
\kappa(\hat{\tau},N)=\frac{\mbox{\rm globule peak area}}
{\mbox{\rm coil peak area}}
\ee
$\kappa(\hat{\tau},N)$ is  the equilibrium constant of the
transition between the  two states
 and it can be deduced from various experimental data\cite{Cantor}.
Such a mapping onto a two-state model relates the theoretical description based
on the knowledge of the configurational statistics and the
% (macroscopic)
experimental observations through a coarse-graining of the configuration space
and not through a thermodynamic limit.
The relevance and the validity of this approach 
will be discussed elsewhere\cite{LV}.
%Let us only underline a few keypoints. 
A  keypoint is that the splitting is independent of the temperature and covers 
the whole configuration space.
%A second point is that for high temperatures ($\tau>0$ finite), the globule region
%has a negligible weight (pure coil state) whereas for 
%low temperatures ($\tau<0$ finite), the coil region
%has a negligible weight (pure globule state).

Estimation of the contribution of each peak to the total area
${\cal I}_c(\hat{\tau},N)$ leads to distinguish two situations.
For $c\leq -1$, the coexistence condition $\kappa(\hat{\tau},N)=1$
writes:
\be 
{-(1+c)\over 2}\ln N
={A^2\over 4B}\hat{\tau}^2+
\ln{\cal C}+c\ln|\hat{\tau}|
\ee
where ${\cal C}$ is some numerical constant.
The coexistence curve $\hat{\tau}_{coex}(N)$ and associated phase diagram
is shown in Figure 2.
It evidences that coexistence in equal proportions of coil and globule phases is
observed only for large enough chains.
% $N>N_{min}\gg N_0$.
The shape of the coexistence curve is similar for any $c\leq -1$;
$\hat{\tau}_{coex}(N)$ behaves as  ${\log N}$ for large $N$.
The coexistence region has a  width 
$\Delta \hat{\tau}\sim 1/\sqrt{\log N}$: it tends to 0 as
 $N\rightarrow\infty$, so that it makes
sense to speak of a ``phase transition'' in the infinite-size
limit\cite{Grosberg}.
As expected,
$\tau_{coex}=\hat{\tau}_{coex}N^{-(1-1/n)}$
tends to 0 as $N\rightarrow\infty$, supporting the  estimate $\theta=J/k_BA$
of the transition temperature.
The shape of the coexistence curve  shows that $N$ is not only
 the size but also a control parameter
ruling the transition: increasing $N$ 
at fixed $\hat{\tau}$ leads into the coil
phase. 

For $-1<c<0$, the coexistence curve is qualitatively different: 
$\hat{\tau}_{coex}(c,N)$ rapidly tends to a finite 
negative asymptotic value. Nevertheless, the thermal behaviour at fixed
%$N>N_{min}(c)$ 
$N$ (enough large) remains unchanged.

In conclusion, for any $c<0$ and 
whatever large is $N$ (and $N>N_0(c)$),  two populations with
 distinct features coexist  in some
range of temperatures;
in this respect, the coil-globule transition thus appears as a first-order
transition, even in the limit as $N\rightarrow\infty$.

\vskip 2mm
Coming back to the unscaled variable $t$, the minimal distance between the two
peaks in the coexistence region
satisfies 
$\Delta t< \hat{x}N^{-(1-1/n)}$, hence tends to 0 as $N$ tends to infinity.
%This follows quite trivially from the size behaviour of the typical globule parameter
%$t_g$ when the globule state rises up:
%$t_g=\hat{t}_gN^{-(1-1/n)}$ tends to 0 as $N$ tends to infinity.
The globule density right at the transition point tends to 0 as 
$\rho_g\sim  N^{-2/5}$.
%From a rough experimental viewpoint, t
This means that in the infinite-size
limit,  
the transition occurs at $\tau=0$ and coil and globule densities coincide
then both to 0.
In this respect, this infinite-size  coil-globule transition
shows some features of a second-order transition, for example the shape of
the mean order parameter $<t>$ with respect  to the reduced temperature $\tau$
has the characteristic shape of a second-order transition,
as shown on Figure 3.
Nevertheless, the transition
occurs through the coexistence of a coil population, whose
statistics is controlled only by the size $N$, and a globule population, whose
statistics is scale-invariant and controlled only by the rescaled temperature
$\hat{\tau}$, as it can be seen on the location and shape of the
corresponding peaks. The first-order
nature of the transition originates in the incompatible scale behaviours
of the two sets of conformations; hence, it is  
 likely to be observed
in  some other coil-globule transitions, for enough large chain sizes, 
as soon as the shape of the 
distribution $P_N(t)$ (infinite-temperature
distribution, describing the purely entropic contribution),
gives enough weight to the coil
region (here for $c<0$).
 A striking signature is the behaviour of the densities
along the coexistence curve:
\be\lim_{N\rightarrow\infty}\rho_g/\rho_c=
(\hat{t}_g/\hat{t}_c)^{4/5}=\infty\ee
showing that in fact, the physical difference between the 
two phases increases with the size $N$.
This first-order nature can be missed in small size but it should appear in
the infinite-size description.
Moreover, it plays a crucial role in the scaling behaviour of the moments.
The first moment writes:
\be
<\hat{t}>\;={{\cal I}_{1+c}(N, \hat{\tau})\over {\cal I}_c(N, \hat{\tau})}
\ee
where 
\be{\cal I}_{1+c}=\int_0^{\infty}
\hat{t}^{c+1}e^{-A\hat{\tau}\hat{t}-B\hat{t}^n}
e^{-A^{\prime}N^{-q(1-1/n)}\hat{t}^{-q}}d\hat{t}
 \ee
Whereas the scaling behaviour of ${\cal I}_{1+c}(N, \hat{\tau})$
depends only on the sign of $\hat{\tau}$, the scaling behaviour of
${\cal I}_c(N, \hat{\tau})$ differs on each side of the coexistence curve  
$\hat{\tau}_{coex}(c,N)$ represented on Figure 2. Indeed, the contribution
of the globule peak  to
${\cal I}_c$ is overwhelming on the left-side (i.e. below) of the coexistence curve
whereas the contribution
of the coil peak is overwhelming on its right-side (i.e. above):
\be
{\cal I}_{c}\propto\left\{
\begin{array}{ll}
N^{-(c+1)(1-1/n)}&\mbox{\rm above}\\
&\\
\left({-A\hat{\tau}\over nB}\right)^{c\over n-1}
e^{(n-1)B(-A\hat{\tau}/nB)^{n\over n-1}}&\mbox{\rm below}
\end{array}
\right.
\ee 
An intermediate scaling region 
$\hat{\tau}_{coex}(c,N)<\hat{\tau}<0$ thus appears, where an anomalous
tricritical scaling is observed. A detailed presentation of the various
scaling regimes following from the first-order nature of the transition
will be presented in a following paper\cite{LV2}.

\vskip 2mm
$\hat{P}_N(\hat{\tau},\hat{t})$ actually describes the configurational
statistics of the
 population provided the thermal equilibrium hypothesis
is satisfied.
This is valid as soon as the system never remains trapped in some
region of the configuration space.
A practical criterion is to check that there is no bottleneck between 
the two peaks
hence no prohibitory barrier between the associated sets of configurations;
a rigorous criterion would require to describe the chain dynamics and  is
far beyond the scope of the equilibrium picture sketched here.

Although our model of energy
 is too crude to account for all the details of the actual
coil-globule transition, some experimental results support the given picture.
Yoshikawa et al. indeed observed the  coexistence of coil-like and globule-like
conformations in a dilute solution of DNA segments,
marked all along their length with fluorescent probes\cite{Yosh}.
Such coexistence cannot be accounted for in the standard tricritical picture.

\vskip 2mm
We believe that our example illustrates the different behaviours that may
underlie a  conformational
 transition, as those involved in biological
 functioning. 
 The moral  is two-fold:

\noindent
--- in numerical simulations of conformational transitions, 
the relevant quantity to be studied is the distribution of the order parameter, here
$P_N^{\beta}(t)$.
Analysis  of the moments  is not sufficient to 
reveal the underlying first-order nature of the transition.

\noindent
--- in experimental studies of conformational transitions, the first question
to be answered is whether the coexistence of two distinct populations can be
observed in some range of temperatures; such an occurence justifies
to analyze experimental data within a two-state model.
Single molecule observations, now at hand,  allow a direct determination
of the order parameter distribution; the framework here presented 
provides a guideline for such novel experimental studies.

\vskip 15mm

{\bf Captions}

\vskip 5mm\noindent
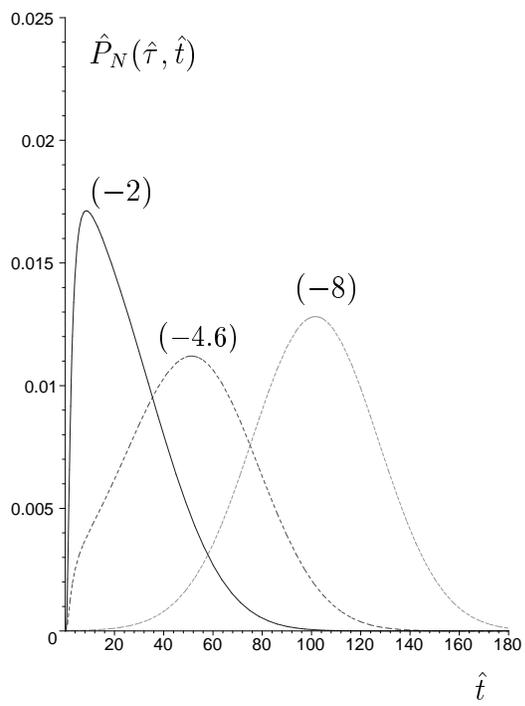
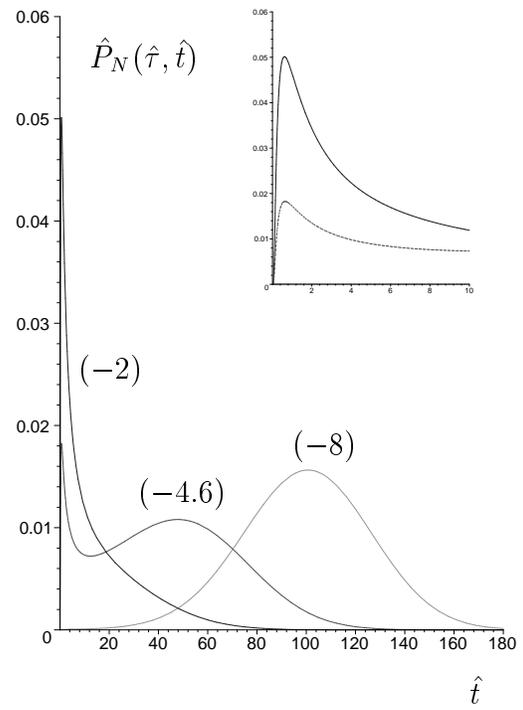
{\bf Figure 1:} 

{\bf (a)} Continuous transition for small sizes $N<N_0$. Here $N=20$,
$c=-1.13$ and $N_0=45$.
The evolution of the shape of the graph of $\hat{P}_N(\hat{\tau},\hat{t})$
as the rescaled temperature $\hat{\tau}$ decreases
(respectively $\hat{\tau}=-2$, $\hat{\tau}=-4.6$ and 
$\hat{\tau}=-8$) clearly indicates a continuous transition in which the
characteristics of a single population evolves smoothly with $\hat{\tau}$.

{\bf (b)} Evidence  of a first-order transition
on the shape of 
$\hat{P}_N(\hat{\tau},\hat{t})$, plotted with respect to
the rescaled variable $\hat{t}$ at fixed $N$
and for various $\hat{\tau}$.
Here $N=2000$ and $c=-1.13$.
Curve  $\hat{\tau}=-2$: only a coil peak is present.  
Curve  $\hat{\tau}=-8$:  only a globule peak is present.
Curve  $\hat{\tau}=-4.86$: two peaks exist, leading to the coexistence
of two distinct populations in dilute solution.
The inset shows an enlarged view of the coil region ($\hat{t}<10$).

\vskip 5mm\noindent
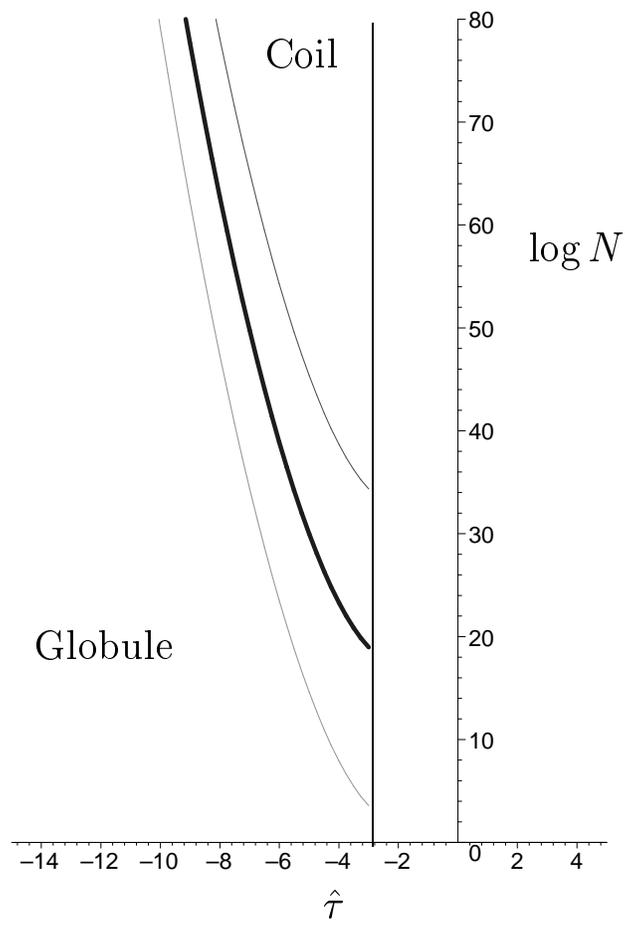
{\bf Figure 2:}  Phase diagram in
$(\hat{\tau}, \log N)$-space
 for $c\leq -1$ (here $c=-1.13$). The vertical straight line
$\hat{\tau}=\hat{\tau}_g$ bounds above the domain of temperatures
where well-identified globule state (a globule peak) exists.
The bold curve corresponds to the coexistence
in equal proportions of coil and globule populations
($\kappa=1$ in equation (\ref{kappa})); it 
 behaves as $\sqrt{\log N}$ for large $N$ .
The neighbouring curves bound the coexistence region ($\kappa=10$ on the
globule side and $\kappa=0.1$ on the coil side)
of width  $\Delta \hat{\tau}(N)\sim 1/\sqrt{\log N}$
for large $N$.
A first-order transition occurs when $\hat{\tau}$ increases at fixed $N$
or when $N$ increases at fixed
$\hat{\tau}<\hat{\tau}_g$.

\vskip 5mm\noindent
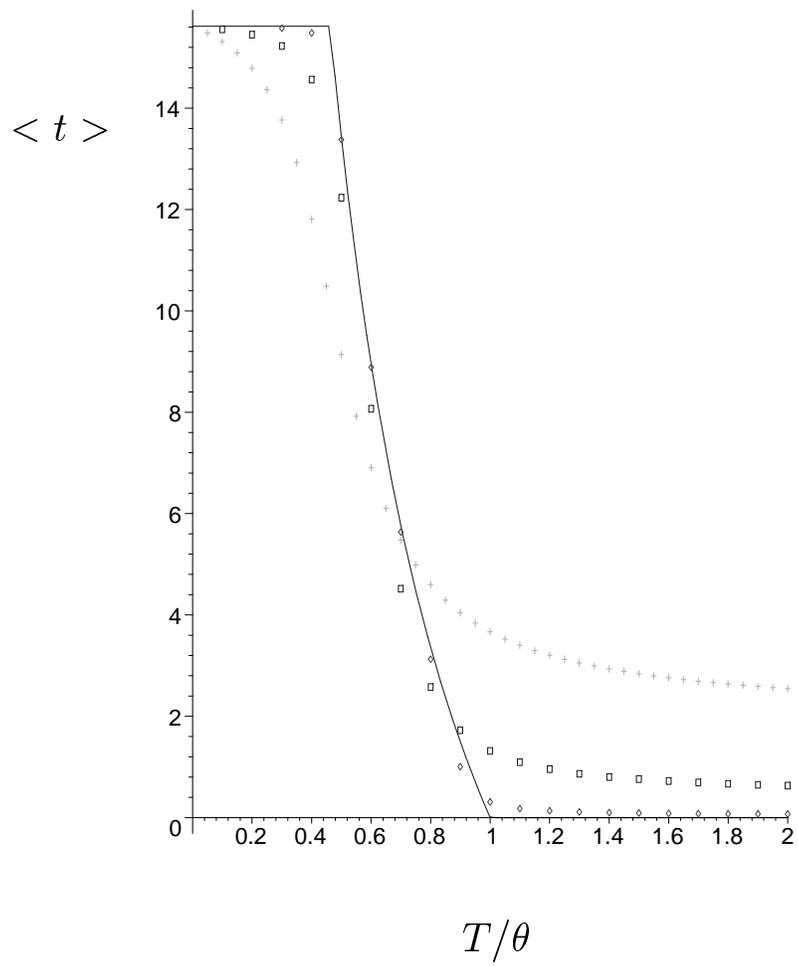
{\bf Figure 3:} Plot of the average order parameter $<t>$ with
respect to the reduced temperature $\tau$ for $N=20$ ($+$),
$N=100$ ({\small $\Box$}) and $N=1000$ ($\diamond$);
in the limit as $N\rightarrow\infty$, the curve exhibits the typical shape of a
second-order transition.

\end{document}